\documentclass[%
reprint,
superscriptaddress,
showpacs,preprintnumbers,
 amsmath,amssymb,
 aps,
pra,
]{revtex4-1}

\usepackage{graphicx}% Include figure files
\usepackage{dcolumn}% Align table columns on decimal point
\usepackage{bm}% bold math
\usepackage{xcolor}
\begin{document}

\preprint{APS/123-QED}

\title{ Experimental distillation of bi-partite polarization entanglement using polarizing Mach-Zehnder interferometers }% Force line breaks with \\

\author{Chithrabhanu Perumangatt}
 \affiliation{Centre for Quantum Technologies, National University of Singapore.}%Lines break automatically or can be forced with \\
 \author{Tang Zong Sheng}
 \affiliation{Centre for Quantum Technologies, National University of Singapore.}%Lines break automatically or can be forced with \\
\author{Alexander Ling}%
\affiliation{Centre for Quantum Technologies, National University of Singapore.}%Lines break automatically or can be forced with \\
\affiliation{Physics Department, National University of Singapore, 2 Science Drive 3, S117542}

\date{\today}% It is always \today, today,

\begin{abstract}
Entanglement distillation is the process of concentrating entanglement from a given quantum state. We present a technique for distillation of bi-partite polarization entanglement using interferometry. This technique can be optimized to extract maximal entanglement from any pure or mixed entangled state. A model for this method is presented and in particular we present experimental results for pure states when using polarizing Mach-Zehnder interferometers. These experimentally distilled states always demonstrate an increased violation of Bell's inequality.
% An article usually includes an abstract, a concise summary of the work
% covered at length in the main body of the article. 
% \begin{description}
% Secondary publications and information retrieval purposes.
% \item[PACS numbers]
% May be entered using the \verb+\pacs{#1}+ command.
% You may use the \texttt{description} environment to structure your abstract;
% use the optional argument of the \verb+\item+ command to give the category of each item. 
% \end{description}
\end{abstract}

\maketitle
% * <chithrabhanup@gmail.com> 2018-02-22T09:30:11.341Z:
%
% ^.
%\tableofcontents

%\section{Introduction}

Entanglement is an important resource for many quantum information protocols \cite{nielsen_chuang_2010, PhysRevLett.67.661, PhysRevLett.70.1895, PhysRevLett.69.2881, PhysRevLett.85.2010, Jozsa2011, PhysRevLett.80.3891, PhysRevA.89.062320}. For convenience, many proof-of-principle implementations of these protocols were performed using polarization entangled photon pairs \cite{ PhysRevLett.84.4729, PhysRevLett.76.4656,bouwmeester1997experimental, Poppe:04}. For better performance of these protocols, high quality entangled states are needed. Sometimes it is necessary to increase the entanglement quality of a given state, either due to imperfections at the source, during transmission or a combination of both effects. 

Entanglement distillation (or concentration) extracts maximal entanglement from a non-maximally entangled state \cite{PhysRevA.53.2046,PhysRevA.64.014301,PhysRevA.64.012304,PhysRevLett.90.207901}. 
It is important to note that distillation does not increase the purity of the given state; this is achieved via a related method called entanglement purification \cite{PhysRevLett.76.722,pan2001entanglement,pan2003experimental}. 
Experimental demonstration of distillation for general bi-partite entangled states have been performed with photon pairs with a probabilistic $C_{NOT}$ gate \cite{PhysRevLett.90.207901,yamamoto2003experimental}. 
One criteria in the above experiments was to act on a general state with no a priori knowledge of what the state was.
However, in many cases, it is possible to have some knowledge of the given state that is to be concentrated.
In this scenario, distillation can be performed more efficiently with local filtering and classical communication\cite{PhysRevA.64.010101, zhang2004entanglement}. 
This has been successfully shown for polarization entangled photon pairs using an experimental technique that relied on polarization dependent reflectors \cite{kwiat2001experimental,PhysRevLett.96.220505,PhysRevLett.92.133601}. 
We note, however, that these previous methods were not tunable across a range of input states; for maximal entanglement distillation the reflectors had to be replaced accordingly for different input states.

This paper reports on a method for the controlled distillation of any (known) bi-partite non-maximally entangled state in a continuously tunable manner. This method employs a Mach-Zehnder interferometer that uses polarizing beamsplitters (instead of the usual non-polarizing devices). Half-wave plates located within the arms of the interferometer can be adjusted, effectively tuning the distillation to increase the amount of entanglement for any given non-maximally entangled state. The increased presence of entanglement is demonstrated by performing the standard Clauser-Horne-Shimony-Holt (CHSH) \cite{PhysRevLett.23.880} entanglement witness, on the given photon pair state before, and after distillation.

For brevity in the following discussion, we adopt the expression below for a pure, non-maximally entangled bi-partite state:
\begin{equation}
|\Phi\rangle =  \alpha\vert HH\rangle+ \beta\vert VV\rangle \label{1}
\end{equation}
with $\vert \alpha\vert^2+\vert\beta\vert^2=1 $. This is a state where the polarization of the photons are aligned. The method introduced below also works for the bi-partite state with anti-aligned polarization, if a half-wave plate is used to convert the polarizations to the aligned state \cite{kok2010introduction} before distillation.
The state in equation~\ref{1} reduces to the Bell state $|\Phi^+\rangle$ when $\beta=\alpha =\frac{1}{\sqrt{2}}$. 
The degree of entanglement for this pure state varies with the relative values of $\alpha$ and $\beta$. 
It is possible to re-parametrize equation \ref{1} as 
\begin{equation}
|\Phi_{\epsilon}\rangle =  \sqrt{\epsilon}\vert HH\rangle+ e^{i\phi}\sqrt{1-\epsilon}\vert VV\rangle \label{2}
\end{equation}
where $\epsilon =\vert\alpha\vert^2$ and $\phi$ is the relative phase between the horizontal and vertical photon components. 

   \begin{figure}[t]
    \centering
        \includegraphics{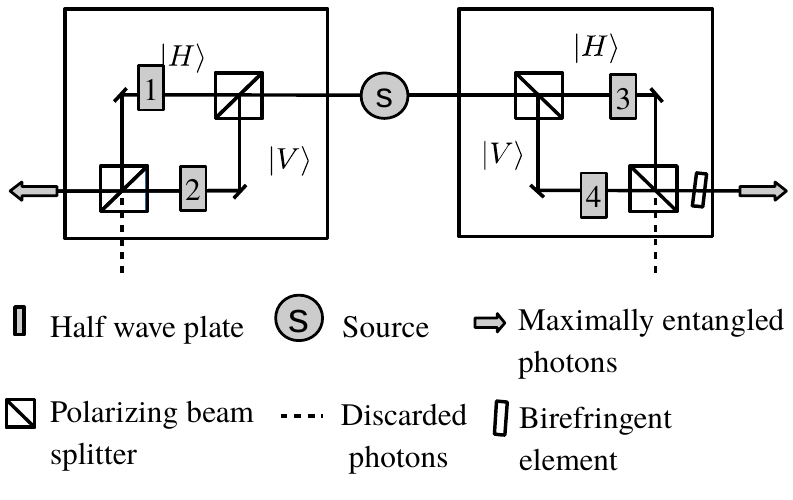}
        
        \caption{Experimental schematic for performing bi-partite entanglement distillation using two polarizing Mach-Zehnder interferometers. Entangled photon pairs from a source are separated; each photon in the pair traverses an interferometer. By appropriately setting the half-wave plates, unwanted photons are discarded, leaving maximally entangled photons to exit the interferometer in the desired output port. }
\label{fig:0}
\end{figure}
        
A maximally polarization-entangled state can be obtained from the description in equation \ref{2} by using polarizing Mach-Zehnder interferometers. 
The schematic for such a distillation method is given in Fig.~\ref{fig:0}.
Photon pairs from a source are separated and each photon is filtered independently through a Mach-Zehnder interferometer.
Within the interferometer, photons are projected into the $\vert H\rangle$ or $\vert V\rangle $ states by the first polarizing beam splitter (PBS), before recombining at the second PBS. 
A half-wave plate (HWP) is placed within each arm of the interferometer.
The appropriate setting of the HWP allows the collection of the maximally entangled photons in one output port of the interferometer; the rejected photons will exit in the other output port.

\begin{figure*}[t]
\centering
\includegraphics{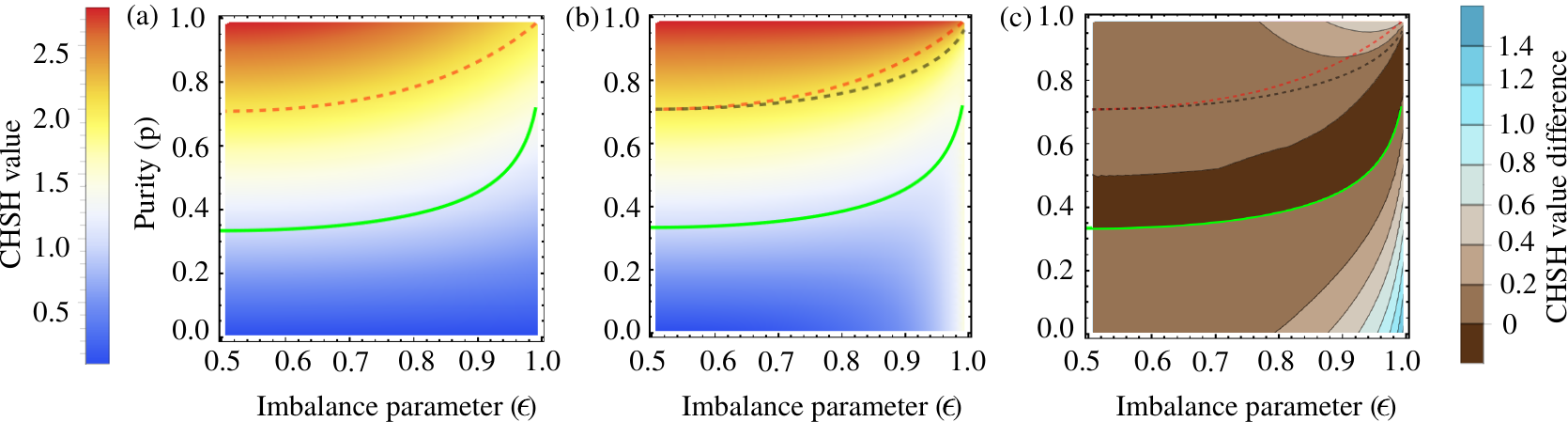}
\caption{ (Color online) Theoretical plots for the CHSH values that is attainable for bi-partite states before (Eq.~\ref{3}) and after (Eq.~\ref{4}) the distillation. This CHSH test is used as an indicator of the non-local quantum correlation between the two particles. The green line indicates the boundary between entangled and separable states. (a) The initial CHSH value for all the states before distillation. The red dashed line is the boundary where the CHSH value is $\geq 2$. (b) The CHSH values after distillation. The black dashed line show the region whose CHSH value exceeded 2 after distillation. (c) Taking the difference between panels (a) and (b) reveal that distillation improves the bi-partite correlation best for states that were initially heavily imbalanced, as expected. There exists a region where the distilled states show worse CHSH values, but states with these starting values are typically of no interest in quantum communications. (Color online).}
%\textcolor{red}{Try different heatmap colors to better show the improvement in correlations. Currently, these maps are very low contast. You need to label also the color bar. The font size also needs to be changed. You cannot import figures directly from Mathematica; it must be edited for a suitable font size.} }
\label{fig:2}
\end{figure*}

Consider Fig.~\ref{fig:0} and the case where the imbalance is $\epsilon > 1/2 $.
To distill a maximally entangled state, the wave plates HWP1 and HWP3 should be adjusted such that $\vert HH\rangle \rightarrow \sqrt{\frac{1-\epsilon}{\epsilon}}\vert HH\rangle$. The wave plates HWP2 and HWP4 are set such that they do not rotate the $\vert VV\rangle$ state.
The bi-partite state becomes $\sqrt{2(1-\epsilon)}\vert\Phi^+\rangle$.  
With these settings, the maximally entangled state is obtained with an efficiency of $2(1-\epsilon)$.
Similarly for the case where the $\epsilon < 1/2 $ we can rotate HWP2 and HWP4 for transforming $\vert VV\rangle \rightarrow \sqrt{\frac{\epsilon}{1-\epsilon}}\vert VV\rangle$, while leaving the $\vert HH\rangle$ state alone.
This causes the bi-partite state to become $\sqrt{2\epsilon}\vert\Phi^+\rangle$, which is obtained with an efficiency of $2\epsilon$. 
When there is perfect balance, $\epsilon=0.5$, the half-wave plate settings are set for no rotation as no distillation is necessary.
The relative phase $\phi$ can be compensated using an additional birefringent element; in Fig.~\ref{fig:0} this is placed after the interferometer on the right.        

We can extend the discussion above to a mixed bi-partite state \cite{PhysRevA.40.4277} with white noise by using density matrix formulation, and introducing the additional parameter of purity (p):
\begin{equation}
\rho = p\vert\Phi_{\epsilon}\rangle\langle\Phi_{\epsilon}\vert+\frac{1-p}{4} I. \label{3}
\end{equation}
Using equation \ref{2}, it is possible to analyze the performance of the polarizing Mach-Zehnder distiller when applied to the general class of bi-partite states.

When the photon pairs described by the state $\rho$ pass through the interferometers, they experience path dependent polarization filtering that balances the horizontal and vertical components. 
The normalized state after post-selection, $ \rho_d^n$, is
\begin{align}
\rho_d^n=\rho_d/\text{Tr}\left(\rho_\text{d}\right), \label{4}
\end{align} 
where
\begin{align}
\rho_d= p\vert\Phi^+\rangle\langle\Phi^+\vert+\frac{1-p}{4} \rho_{\text{noise}},
\end{align} 
and
\begin{align}
\rho_\text{noise}=&\frac{1-\epsilon}{\epsilon}\vert HH\rangle\langle HH\vert+\vert VV\rangle\langle VV\vert +\notag\\&
\sqrt{\frac{1-\epsilon}{\epsilon}}\left(\vert HV\rangle\langle HV\vert+\vert VH\rangle\langle VH\vert\right).\label{3a}
\end{align} 

The term for white noise in  Eq. \ref{3} has transformed to colored noise (Eq. \ref{3a}) due to the distillation. Furthermore, due to the renormalization, the efficiency of extracting the maximally entangled state with respect to the noise is a function of $\epsilon$ and $p$.

The results of interferometric distillation on the input states are shown in Fig.~\ref{fig:2} where the starting states (before distillation) are parametrized by their balance ($\epsilon$) and purity (p). The behavior of CHSH values is symmetric about $\epsilon=0.5$;  hence the plot only shows the value where $0.5<\epsilon<1$. %Fig.~\ref{fig:2} uses a mirror projection about $\epsilon=0.5$; hence the plot only shows the value where $0.5<\epsilon<1$. 
No distillation occurs for $\epsilon=0.5$,~ (where $|\Phi_{\epsilon}\rangle = |\Phi^+\rangle)  $ or $\epsilon=1$,~ (where $|\Phi_{\epsilon}\rangle = |HH\rangle)$.
\begin{figure} [b]
            \includegraphics{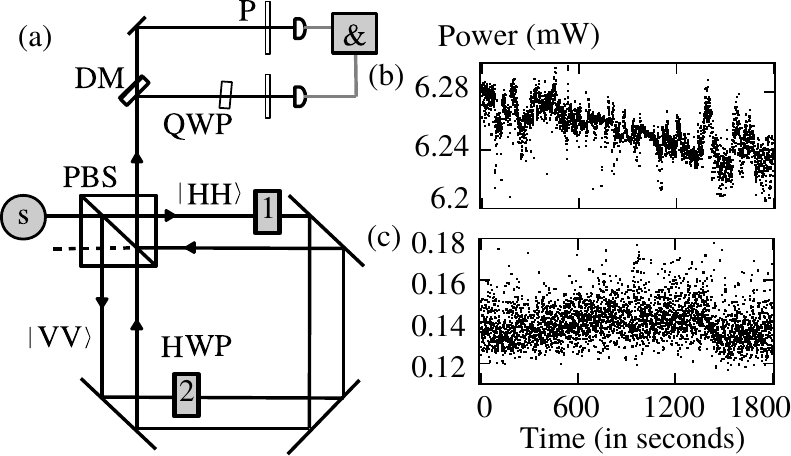}
        \caption{(a)Experimental setup for the distillation using a folded Mach-Zehnder interferometer. A photon pair source (S) capable of producing non-maximally entangled photons with different $\epsilon$ is used. The dotted lines are the untangled photons filtered out by the interferometer. Two polarizers (P) are used to measure the polarization correlations that contribute to the CHSH parameter. A quarter wave plate oriented at $0$ degree introduces a phase between horizontal and vertical component depending upon its tilt angle. This phase can be used to compensate any phase offset created either by the source or the interferometer. A dichroic mirror (DM) separates the distilled signal and idler photons.  Figures (b) and (c) shows the stability of the constructive and destructive interference created by the folded Mach- Zehnder interferometer. The small drift in the constructive interference output (approximately 1\%) is attributed to the shift in the input laser power. However, there is no phase instability as the constructive/destructive interference contrast remains the same over the course of the experiment.} 
        \label{fig:5}
        \end{figure}

In Fig.~\ref{fig:2}(a), the maximal attainable CHSH value for a given input state is provided; the maximally entangled state is in the top left-hand corner of this plot.
Fig.~\ref{fig:2}(b) shows the CHSH value attained after the distillation. Notably, the region of states that can now achieve a CHSH value greater than 2 has been increased as shown by the gap between the black and red lines. In particular, all the pure states (where $p$ is close to 1 at the top of the graph) are able to achieve CHSH values close to $2\sqrt{2}$.
For greater clarity, a difference of the CHSH values before and after distillation are provided in Fig.~\ref{fig:2}(c) revealing regions with the largest increase in correlation quality. 
Not surprisingly, these are mostly in the region where the states were originally very heavily imbalanced. 

It is notable that the region showing the largest increase in correlation quality is for the separable states near the bottom right-hand corner; however, the improvement is not sufficient to exhibit a violation of the CHSH inequality as expected for a non-purification protocol \cite{PhysRevA.53.2046}. There is also a region where the CHSH value decreases after distillation; this is caused by the renormalization process which reduces the ratio of pure state with respect to the noise for highly imbalanced initial states. Finally, distillation does not move the separability bound as indicated by the green line.

%\textcolor{red}{``However, there is a region where the CHSH value decreases after distillation, this is caused by the renormalization of state will reduce the ratio of pure state with respect to the noise." This statement that I found in the text is not clear in the graphs, and needs further elaboration. I would not talk about this unless it is very important.}
%\textcolor{green}{ You can observe some negetive values in the CHSH value difference plots. However, we can skip mentioning the same. They are not so important as the states are alrealy lie below the separability bound. }
%\textcolor{red}{How is it even possible to see the negative values when you use the same heatmap for panels (a), (b) and (c)? Also the contrast is so poor, it conveys no information. Make sure you change the heatmap color. You may want to use a log scale on the heatmaps to show better contrast.}

The polarizing interferometer scheme was then tested experimentally using pure, bi-partite states as these are straightforward to obtain, and of greatest interest in the context of quantum communication \cite{gisin2007quantum, ursin2007entanglement, ma2012quantum}

The bi-partite states were implemented using Spontaneous Parametric Downconversion (SPDC) where photon pairs are generated when a pump photon undergoes frequency conversion within a material exhibiting $\chi^2$ optical nonlinearity.
We have utilized a type-1, critically phase matched SPDC source that produced collinear, non-degenerate photon pairs from cascaded $\beta$-Barium Borate crystals whose optical axes were crossed \cite{doi:10.1063/1.2924280}. 
The use of non-degenerate phase matching enables the signal and idler photons making up the photon pair to be separated easily with a dichroic element.
Type-1 phase matching enabled us to produce photon pairs whose polarization state was described by equation~\ref{2}.
The photon pairs generated from the first crystal have a polarization which is orthogonal when compared to photon pairs generated in the second crystal. 
By appropriately overlapping the SPDC emission the photon pairs become polarization entangled.

The balance, $\epsilon$, between the horizontal and vertical components can be adjusted by placing a half-wave plate that acts on the pump polarization before the SPDC process.
Typically, to generate the maximally entangled $\vert \Phi^+ \rangle$ state, the wave plate is oriented such that the pump polarization will perform SPDC with equal probability within each of the $\beta$-Barium Borate crystals.
By setting the wave plate angle, it is possible to imbalance the SPDC rate from each crystal thereby creating the non-maximally entangled state. 
\begin{figure*}[t]
    \centering
         \includegraphics{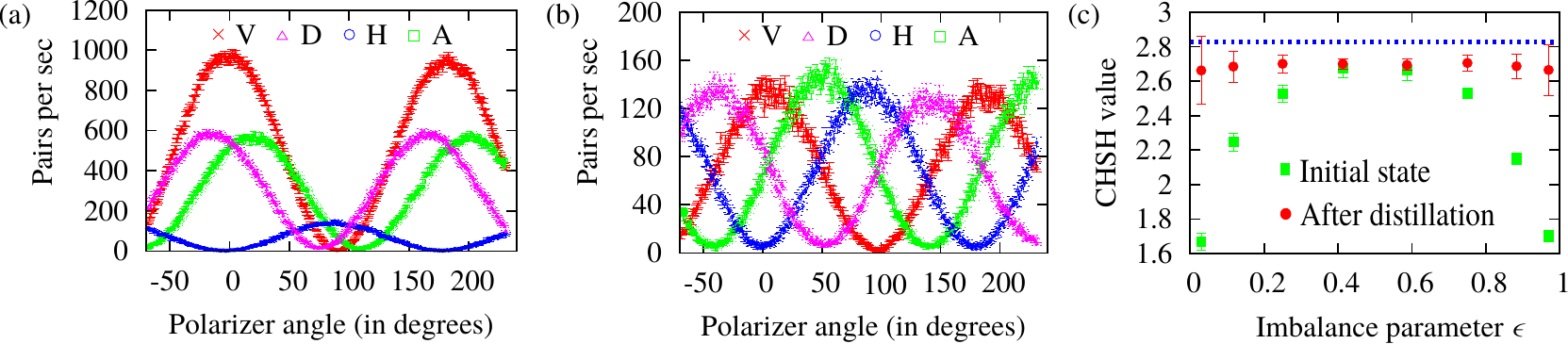}
        \caption{ (Color online)(a) Two photon polarization correlation for a non-maximally entangled state with $\epsilon=0.11$. V, D, H and A correspond to the vertical, diagonal, horizontal and anti-diagonal polarization projections on the signal photon. (b) Polarization correlation curves for the distilled state. Visibility of the curves are unchanged during the distillation. The diagonal basis visibility is observed to be 94\%. (c) CHSH value for different non-maximally entangled states before and after distillation when measured in mutually unbiased bases. Blue dotted line is the upper bound ($2\sqrt{2}$) for the CHSH value. The imbalance $\epsilon = cos^2(2\theta)$, where $\theta$  correspond to the angle of the HWP that rotated the pump polarization. When the imbalance parameter have extreme values, distillation procedure filters out most of the photons from the source resulting in a low signal to noise ratio, increasing the standard deviation of the measured CHSH values.
}         \label{fig:6}
         \end{figure*}
One challenge when working with bulk interferometers is stability, and often these instruments are actively stabilized using a reference laser.
To improve the stability, and do away with active stabilization, a folded Mach-Zehnder inteferometer was implemented as shown in Fig.~\ref{fig:5}(a). 
The folded geometry enables all the light paths to be defined by the same reflecting and dividing elements, improving the stability.
The setup exhibits sufficient passive stability of at least 24 hours (under general laboratory conditions), determined by observing the bright/dark (constructive/ destructive interference) output ports of the interferometer when pumped with a coherent 808 nm laser beam.
{Fig.~\ref{fig:5}(b) and ~\ref{fig:5}(c) show the power at the constructive/destructive interference output ports of the interferometer for a sample period of approximately 30 minutes. 
For an  input power of 6.5 mW, constructive (destructive) interference is observed to be  $6.25\pm 0.016 mW$ ($0.139\pm0.009 mW $).
Stability over this duration was sufficient for the purpose of the experiment.

Following Fig.~\ref{fig:5}(a), the polarizing beam splitter (PBS) will separate the $\vert HH\rangle$ and $\vert VV\rangle$ photon pairs.
Half-wave plates are inserted into each arm.
The optical beams recombine at the other edge of the PBS. 
If both the half-wave plates fast axes are oriented at $0$ degree (this is known as the neutral position), no distillation takes place, and all photons emerge through one port of the PBS. 
To carry out distillation the wave-plates are adjusted appropriately so that the unentangled photons get filtered out in the other output port. 
For performing the CHSH test, the signal and idler photons are separated using a dichroic element before being interrogated individually with polarization analyzers.
A birefringent element (made by tilting a quarter-wave plate) is placed in the signal arm to compensate for any additional phase introduced by the interferometer. This can also be used to compensate any inherent relative phase $\phi$ given in Eq.~\ref{2}.

In this experiment, 8 different non-maximally entangled pure states were generated  over a range of $\epsilon$ values (see Fig.~\ref{fig:6}(c)). 
As stated previously, the imbalanced state is $\vert \Phi\rangle = cos(2\theta)\vert HH\rangle + sin(2\theta) \vert VV\rangle$, where $\theta$ is the setting of the half-wave plate acting on the pump beam of the SPDC source.
Therefore, $\epsilon = cos^2(\theta)$.  
For $0<\theta<\pi/8$, where the horizontal component has a higher amplitude than the vertical component, the half-wave plate labeled 1 (in Fig.~3) is rotated by $\frac{1}{2}cos^{-1}(\sqrt{tan(2\theta)})$ for achieving optimal distillation. 
Similarly for $\pi/8<\theta<\pi/4$, the half-wave plate labeled 2 is rotated through an angle $\frac{1}{2}cos^{-1}(\sqrt{cot(2\theta)})$ to remove the un-entangled vertically polarized components. 

It is instructive to observe how the detected photon pair rate and the entanglement quality changes through distillation by following one of the test states as it passes through the interferometer.
Consider the highly unbalanced non-maximally entangled state where ($\epsilon=0.11$).
The polarization correlation measurements for this starting state in two different polarization bases are given in Fig.~\ref{fig:6}(a). 
The imbalance in the vertical and horizontal polarizations are clearly visible from the amplitude of the curves. 
Furthermore, the correlation curves in the diagonal/anti-diagonal basis are only slightly displaced from the curves in the vertical/horizontal basis due to the heavy imbalance.

This highly imbalanced input state can be distilled by leaving the wave plate in the transmitted arm of the interferometer in the neutral position, while adjusting the wave plate in the reflected arm by 34.71 degrees. The excess vertical polarization component is filtered out by the PBS (dotted lines in Fig.~\ref{fig:5}(a)).  
After this adjustment, the correlation curves of the distilled state exhibit much better balance in amplitude and position of the extrema as shown in Fig.~\ref{fig:6}(b). The plot in Fig.~\ref{fig:5}(b) exhibits the characteristic polarization correlation expected from a $|\Phi^+\rangle$ Bell state. As the imbalance of the initial state is large, a majority of input photons are filtered out in the distillation procedure leaving the relevant maximally entangled photons.

As a quantitative measure of non-locality and certifiable entanglement \cite{PhysRevA.88.052105}, the CHSH correlation measurement was carried out for the different input and distilled states by projecting the two photon state into mutually unbiased polarization states. The settings used for the measurement of CHSH value, projection angles of the polarizers, for the state before and after the distillation are kept  the same.
%The polarizer in the signal arm is oriented at angles $0^{\circ}, 45^{\circ}, 90^{\circ} $  and $135^{\circ}$ while the polarizer in the idler arm is oriented at $22.5^{\circ}, 57.5^{\circ}, 112,5^{\circ} $  and $157.5^{\circ}$. From the 16 measurement outcomes, CHSH values are calculated.
As shown in Fig.~\ref{fig:6}(c), the CHSH value for the non-maximally entangled (green squares) state is a function of the imbalance parameter $\epsilon$. Where the imbalance is higher, the CHSH value has a reduced value indicating the presence of reduced entanglement. The CHSH value after the distillation (red dots) is well-balanced and always corresponds to a highly entangled state. 

%For every value of $ \epsilon$ (except 0,1 and 0.5) the distillation increases the CHSH value. This is only true in the case of pure state. When the state is mixed, as shown in Fig.~\ref{fig:2}, the CHSH value for the distilled state is also a function of initial imbalance parameter. However for a large class of state distillation will increase the CHSH value.  The increase in the CHSH value is higher when the initial states are heavily imbalanced. 

The reported interferometric method offers good tunability for the distillation of bi-partite polarization entanglement as it can be applied to a large range of inputs with different imbalance parameters. This method can be extended to multi-photon GHZ states \cite{greenberger1990bell, PhysRevLett.117.210502} with interferometers on each photon paths.
The experimentally distilled states show a significant violation of Bell's inequality even for highly imbalanced inputs, while using a folded polarizing Mach-Zehnder interferometer significantly improves the stability for the distillation scheme. 
In a two-party quantum communication scenario, two similar interferometers (one in each photon path) can be used to implement the distillation. 
This functionality may be of value when building future quantum communication networks.  
  
\section{Acknowledgements}
This material is based upon work supported by the Air Force Office of
Scientific Research under award number FA2386-17-1-4008. This program is also supported
by the National Research Foundation, Prime Minister's Office of Singapore, and the Ministry of Education, Singapore. The authors like to knowledge Arian Stolk for his inputs in the initial discussions and literature survey. 
\end{document}